\documentclass[12pt]{iopart}
\usepackage{graphicx}
\usepackage{color}
\usepackage{iopams}  
\usepackage{verbatim}
\usepackage{cite}

\begin{document}

\title[]{Kinetic simulation of the sheath dynamics in the intermediate radio-frequency regime}

\author{{M. Shihab}$^{1,4}$, A. T. Elgendy$^1$, I. Korolov$^2$, A. Derzsi$^2$, \\ J. Schulze$^3$,
 D. Eremin$^1$, T. Mussenbrock$^1$, Z. Donk\'o$^2$, \\ and R. P. Brinkmann$^1$}

\address{$^1$ Ruhr-University Bochum, Institute for Theoretical Electrical Engineering, D-44801 Bochum, Germany\\
	     $^2$ Institute for Solid State Physics and Optics, Wigner Research Centre for Physics, Hungarian Academy of Sciences, Budapest, Hungary \\
	      $^3$ Institute for Plasma and Atomic Physics, Ruhr-University Bochum, Germany\\
				$^4$ Department of Physics, Faculty of Science, Tanta University, Tanta 31527, Egypt}

\ead{Mohammed.Shihab@ruhr-uni-bochum.de}

\begin{abstract}
The dynamics of temporally modulated plasma boundary sheaths is studied in the intermediate radio frequency regime where the applied radio frequency and the ion plasma frequency are comparable.
Two kinetic simulation codes are empl\-oyed and their results are compared.
The first code is a realization of the well-known scheme, Particle-In-Cell with  Monte Carlo collisions (PIC/MCC) and simulates the entire discharge,
a planar radio frequency capacitively coupled plasma (RF-CCP) with an additional heating source. The second code is based on the recently published scheme Ensemble-in-Spacetime (EST);
it resolves only the sheath and requires the time resolved voltage across and the ion flux into the sheath as input.
Ion inertia causes a temporal asymmetry (hysteresis) of the sheath charge-voltage relation; also other ion transit time effects are found.
The two codes are in good agreement, both with respect to the spatial and temporal dynamics of the sheath and with respect to the ion energy distributions at the electrodes.
It is concluded that the EST scheme may serve as an efficient post-processor for fluid or global simulations and for measurements:
 It can rapidly and accurately calculate ion distribution functions even when no genuine kinetic information is available.
\end{abstract}

\vspace{2pc}

\maketitle

\baselineskip=20pt

\section{Introduction}

Plasma enhanced surface modification employs the energy of plasma-generated particles.
For the technology computer-aided design of such processes, the calculation
of the flux, energy distribution, and angular distribution of the surface-incident particles is therefore of great importance \cite{Hamaguchi1999,Makabe}.
In theory, this poses no problems, as the foundation  of plasma simulation is sound \cite{Mussenbrock2012,Donko,kim}.
All species of a discharge  -- electrons, ions, neutrals --  \linebreak
can be described kinetically, i.e. by time evolution equations for the six-dimensional distribution functions $f_s(\vec{r},\vec{v})$ in the domain  $V\!\times\!\mathbb{R}^3$ ($V\subset \mathbb{R}^3$ is the spatial domain, i.e. the discharge chamber;
$\mathbb{R}^3$ is the set of all velocities, and $s$ is a species index).  

In practice, however, the problems are considerable. The numerical effort of directly solving kinetic plasma models is definitely deterring, 
at least for the foreseeable future. Mathematical simplifications must be employed, and, therefore, compromises made. 
One option is to reduce the spatial dimension of the problem by assuming a planar or axisymmetric discharge geometry. 
Stochastic solutions of the kinetic equations, based on \linebreak the particle-in-cell approach  with Monte Carlo collisions (PIC/MCC), are then feasible and indeed yield a lot of insight \cite{Birdsall,BirdsallPICMCC}.
However, this approach cannot capture the often complicated geometries of real discharges.

If the latter requirement is an issue, alternatives to kinetic models must be sought. 
One possibility is the fluid approach where particles  
are not represented by distribution functions but only by fluid variables -- typically by the number density and the flux, 
 and, in the case of the electrons,
the temperature. However, often kinetic aspects of the problem are significant and must be retained. This is the realm of ''hybrid schemes''
which combine fluid and kinetic arguments in some clever way, with the intent  to achieve physically sound kinetic information without paying 
the price of a full kinetic simulation. However, the validity of such approaches is always an issue \cite{Kratzer,Salabas2005,Kushner2009}.

Our study is motivated by a particular type of hybrid scheme, namely one that \linebreak couples a fluid description of the plasma
bulk with a kinetic model of the plasma sheath. 
From the following ordering which is typical for processing (where $L$ is the discharge size, \linebreak $l_{\rm ion}$ is the ionization length, $\lambda$ is the elastic ion mean free path, 
and $s$ is the sheath thickness), one can infer that it is solely the plasma bulk which determines the fluxes of the \linebreak energetic species to the wall (fluid quantities), 
while it is solely the sheath that determines the energy and angular distribution functions at the surface 
(kinetic quantities): 
\begin{equation}
 L \approx l_{\rm ion} \gg s \approx \lambda.
\end{equation}

To determine the energy and angular distribution of the surface-incident particles, \linebreak
such a hybrid scheme should, thus, suffice. 
Most simply, it may be implemented as follows: 
A fluid  code (which may be a hybrid code in other respects) is run to its convergence. 
The fields of the final state  (or, in the RF case, of the final period) are transferred  to a Monte-Carlo module,
which simulates the trajectories of a sufficiently large set of test particles and 
calculates the normalized distribution functions at the selected surface. The absolute distribution functions are finally found
as products of the normalized functions with the respective fluxes (to be determined from the fluid code). 

This approach has been implemented into many simulation codes, for example into the codes HPEM and nonPDPsim 
by Kushner \cite{Kushner2009,kushner2005} and co-workers or the commercial codes COMSOL and CFD-ACE \cite{Comsol,ESI}.
The scheme has, however, one essential draw\-back: It is not self-consistent on the kinetic level. To correct this deficiency, 
we have recently  proposed the novel scheme, \textit{Ensemble-in-Spacetime} (EST), which constructs a uniform solution of the 
kinetic equations for the particles and Poisson's equation for the field, i.e., provides a fully self-consistent kinetic simulation 
of the sheath. Its input parameters are still fluid-dynamic: the fluxes of the energetic species, the voltage across the sheath,
the electron temperature, and the composition of the neutral background \cite{shihab2012}.

Of course, this raises a question: How does EST compare to a full kinetic scheme? \linebreak Is the energy  distribution of the ions incident on the surface given by the EST scheme the same as the one calculated from a complete PIC approach? We will show that the answer to this question is positive.
As a test ground, we will study the sheath dynamics in the regime where kinetic effects are most pronounced, namely in the intermediate 
 regime where the applied radio frequency $\omega_{\rm RF}$ is comparable to the ion plasma frequency $\omega_{\rm pi}$ and the inverse
ion sheath transit time $\tau_{\rm i}$. Here, the ions can only partially respond to the time varying field in the sheath,
and interesting effects can be observed like temporal asymmetries and phase shifts between applied voltage and ion energy
   \cite{Kawamura2,Fadlallah,Jacobs2}.

\pagebreak

\section{Set-up and kinetic models}

\begin{figure}
\center
\includegraphics[clip,width=0.8\linewidth]{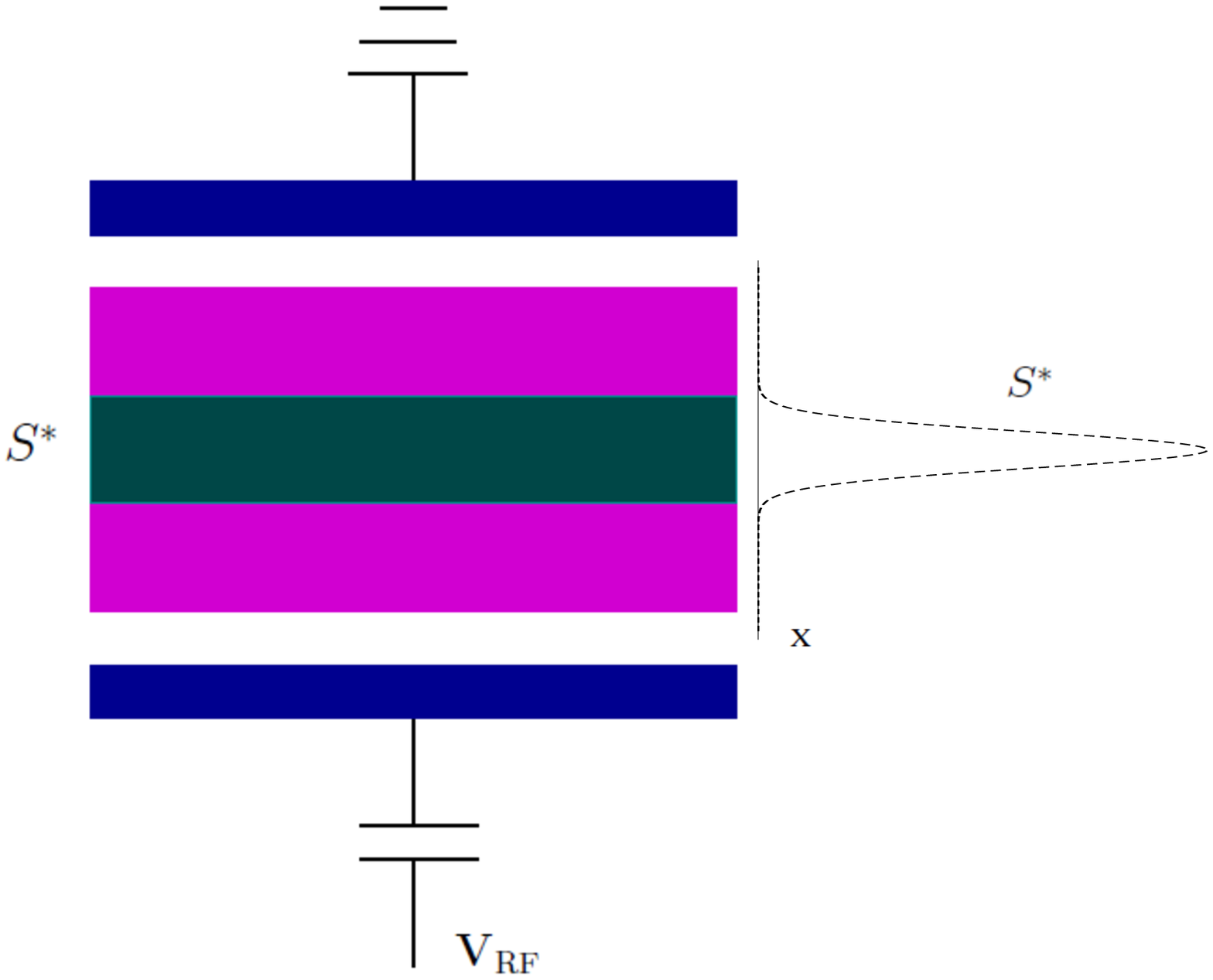}
\caption{Schematic of the case studied in this paper. The discharge is a planar capacitively coupled plasma, run with Argon at $p=1\,$Pa. 
         The electrode gap is~$2\,$cm.\linebreak The shaded area represents the assumed additional Gaussian ionization source $S^\ast$.}
\label{HCCPR}
\end{figure}

The set-up of our study is shown in \Fref{HCCPR}. It is a single frequency capacitively coupled plasma driven at 
a voltage of $V_0 = 100 \,{\rm V}$ and an RF frequency of $\omega_{\rm RF} = 2\pi \times 0.5\,{\rm MHz}$. \linebreak
The electrode gap is $d= 2\,{\rm cm}$, the gas is Argon at a pressure of $p=1\,{\rm Pa}$ and a fixed temperature of $T=300\,{\rm K}$.
All particles that impact the two electrodes are absorbed, the emission of secondary electrons is neglected.
A particular feature of the model is an additional ionization source $S^*$ assumed in the plasma bulk. 
The need for this additional source is purely technical: Single-frequency CCPs with gaps in the cm range \linebreak cannot 
be sustained at the assumed pressure and frequency. An alternative would have been to simulate a double frequency CCP or other hybrid discharge,
but this would have made our results below more difficult to interpret.

\begin{figure}
\center
\includegraphics[clip,width=0.8\linewidth]{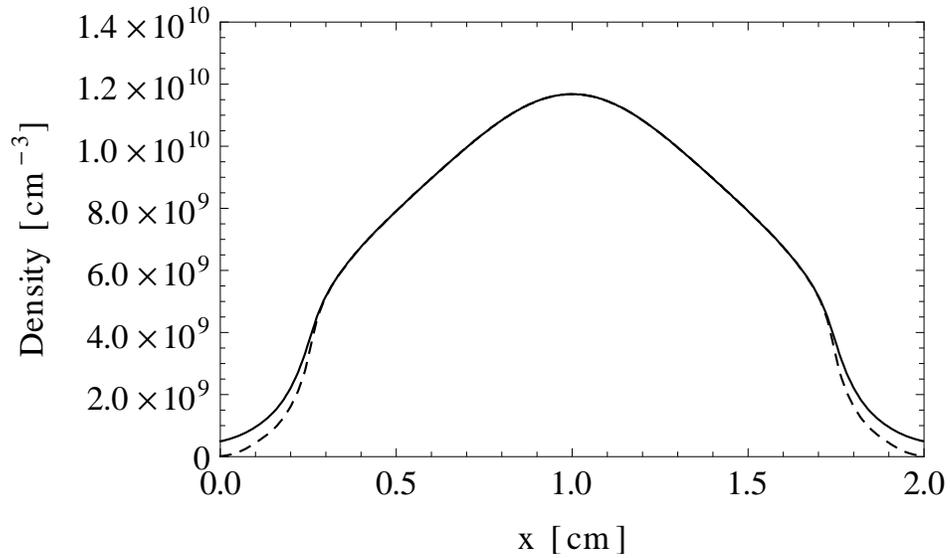}
\caption{Distribution of charged particle densities calculated by the PIC simulation. 
The solid line represents the phase-averaged ion density $\bar {n}_{\rm i}(x)$, the dashed line denotes the phase-averaged 
    electron density ${\bar n}_{\rm e}(x)$.}
\label{density}
\end{figure}

\begin{figure}
\center
\includegraphics[clip,width=0.8\linewidth]{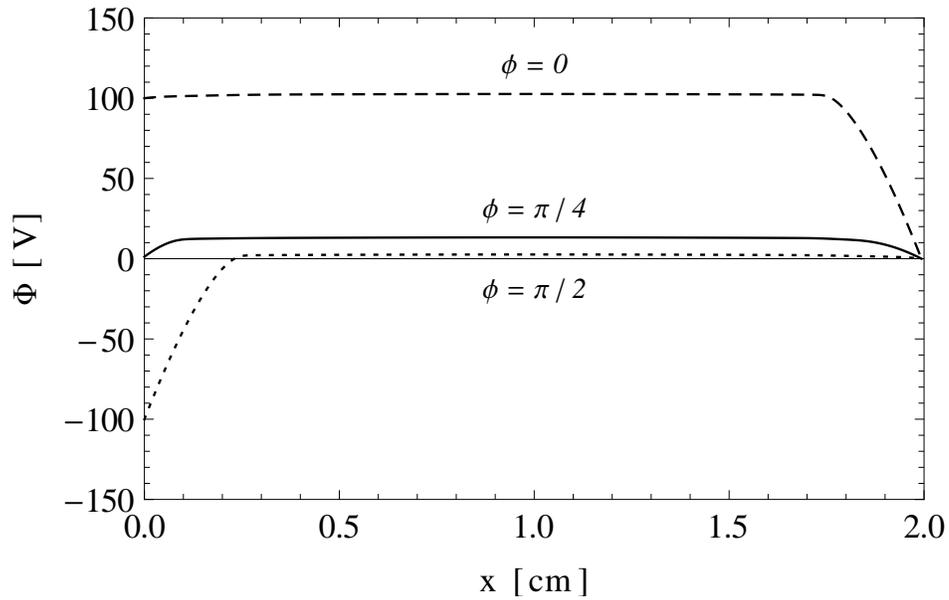}
\caption{The potential along the discharge at different phases within the RF cycle
 (dashed $\phi=0$, solid $\phi=\pi/4$, and dotted $\phi=\pi/2$).}
\label{spatiotemporalPotential}
\end{figure}

The first scheme used in this study is a benchmarked realization of PIC/MCC \cite{Turner}. \linebreak
It is a one-dimensional planar (1d3v) bounded implementation,  which incorporates a Monte Carlo treatment of 
collision processes with the cross sections taken from \cite{Phelps1,Phelps2}.\linebreak  The additional bulk ionization source $S^*$ is 
modeled by electron-ion pair creation in a Gaussian-shaped $2.5\,{\rm mm}$-wide region in the center of the discharge; the average energy of the 
new electrons is $3\,{\rm eV}$. \Fref{density} displays the obtained time averaged density profiles.\linebreak
\figurename{~\ref{spatiotemporalPotential}} shows the time resolved electrical potential.
All expected features are  visible,\linebreak in particular the emergence of a quasineutral bulk with a relatively weak electrical field
and of electron-depleted sheaths with much higher voltage drop. The plasma bulk has \linebreak a positive
potential relative to the electrodes, except for brief moments in the RF cycle, \linebreak where electrons can reach the 
electrodes to balance the ion flux \cite{Schultze2010}.

\begin{figure}
\center
\includegraphics[clip,width=0.8\linewidth]{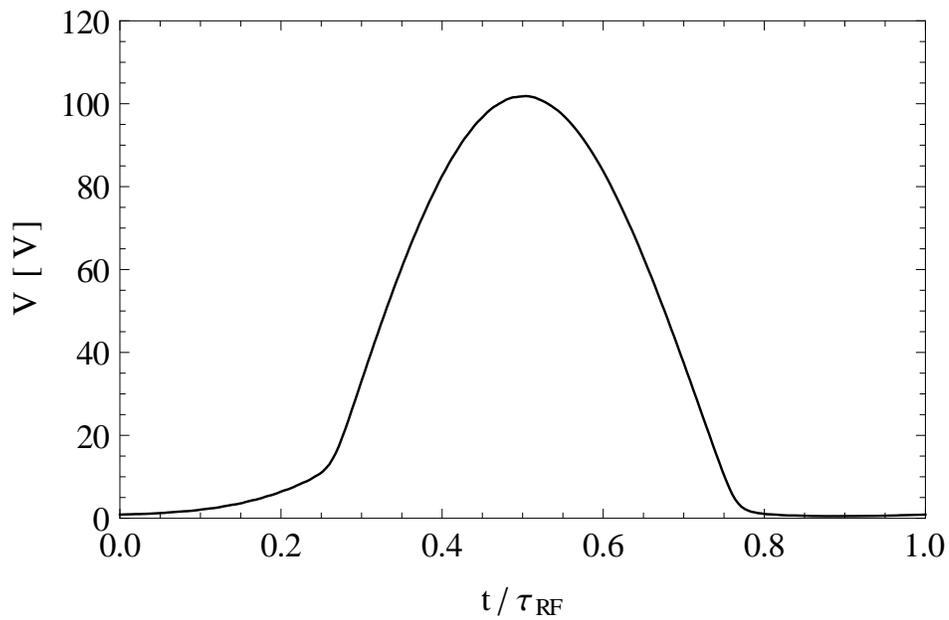}
\caption{The sheath voltage -- the difference between the plasma potential and the electrode potential -- 
    as a function of the normalized time $t/\tau_{\rm RF}$ as calculated by PIC.}
\label{VSH}
\end{figure}

The second algorithm studied in this manuscript, the \textit{Ensemble-in-Spacetime} \linebreak
(EST), was designed as a tool for technology-oriented computer aided design (TCAD).
It~provides a fast, kinetically self-consistent simulation of a DC or 
RF plasma boundary sheath and the resulting ion energy distribution. 
EST differs from PIC in several aspects.\linebreak It does not simulate an entire discharge but only the sheath; the ''operation parameters'' 
-- sheath voltage,  ion fluxes, electron temperature -- must be provided as an input. \linebreak
In our example, the sheath voltage is deduced from the potential as shown in \figurename{~\ref{VSH}}, the ion flux is 
$\Gamma_{\rm i}= 6.25 \times 10^{14}\,\rm{cm}^{-2}\rm{s}^{-1}$,
and the electron temperature is $T_{\rm e}=1.0\, \rm{eV}$.
Like PIC, EST has kinetic equations for the ions which are solved in a stochastic sense. 
However, the electrons are not treated kinetically but assumed to follow a Boltzmann relation with given temperature,
$n_{\rm e} = n_{\rm e0}\exp (e\Phi/T_{\rm e})$. 

\begin{figure}
\center
\includegraphics[clip,width=0.8\linewidth]{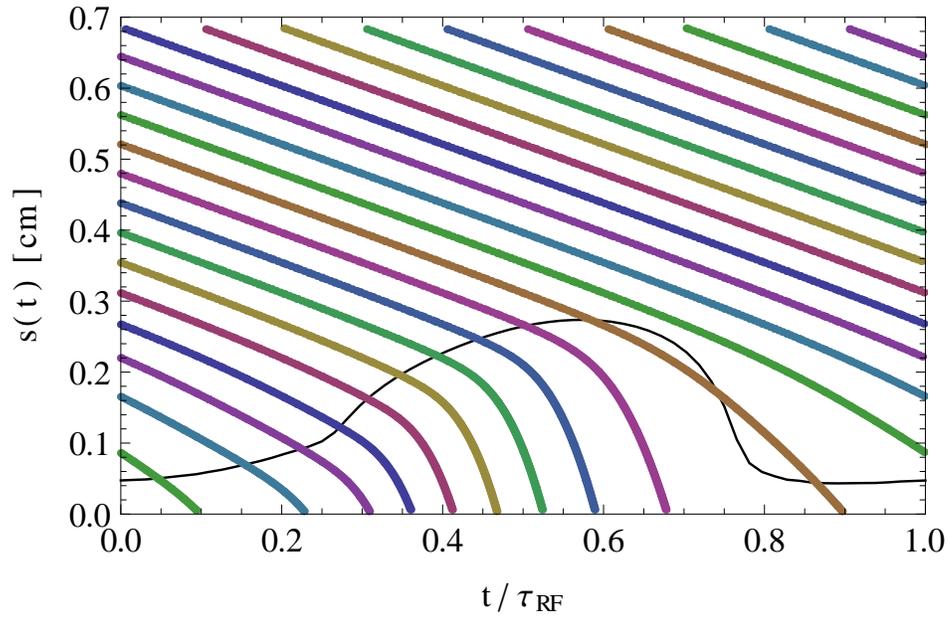}
\caption{The space-time trajectories $x(t)$ of an ensemble of ions (colored lines). The~sheath edge is represented as black line.} 
\label{X}
\end{figure}

Finally, 
EST has a different solution strategy than PIC. It does not follow the transient evolution until it reaches a ''converged'' 
(= periodic) state, but rather seeks within the space of all 
periodic sheath states for a solution of the equations of motion. \linebreak
The central structure of the algorithm is the name-giving \textit{spacetime}, 
a discretized (grid) representation of the domain $[x_{\rm E},x_{\rm B}]\times [0,\tau_{\rm RF}]$.
Here, $x_{\rm E}$ is the location of the electrode, $x_{\rm B}$ is an arbitrary point deep in the plasma bulk, and 
$\tau_{\rm RF}= 2 \pi/\omega_{\rm RF}$ is the RF period. \linebreak The algorithm starts by assigning to each node $(x_k, t_l)$
of the spacetime a potential $\Phi_{kl}$; \linebreak calculated with a fluid sheath model \cite{Brinkmann2007,Brinkmann2009,Brinkmann2011}. 
Then, three modules are iterated:
 \begin{enumerate}
\item
A Monte Carlo module finds the trajectories of the \textit{ensemble}, a large set of test ions. \linebreak
The ions are started at $x_{\rm B}$ with their drift speed and uniform distribution in phase and are followed until they leave the system at $x_{\rm E}$.
Elastic collisions with isotropic and backscattering components with a spatial uniform background gas are performed in the same way as in the PIC code.  
The set of all trajectories -- illustrated in \figurename{~\ref{X}} below -- represents the response of the ions to the field.

\item  
A harmonic analysis module reconstructs from the calculated trajectories and from the prescribed flux 
the ion density $n_{{\rm i}kl}$ for each node $(x_k, t_l)$ of the spacetime grid. 
By construction, it is a periodic quantity.

\item A field module solves for each phase point $t_l$ the Boltzmann-Poisson equation with the calculated ion densities to update the potential.
      The electron density is obtained by Boltzmann's relation, with the electron temperature as specified; the boundary condition is 
			derived from the prescribed sheath voltage.
			
\end{enumerate}

The iteration is terminated when the updates of the potential are sufficiently small. \linebreak
Owing to the stochastic nature of the Monte Carlo module, the algorithm exhibits no convergence in the strict sense, 
but reasonable accuracy (on the $10^{-3}$ level) can typically be achieved in less than five iterations.

\pagebreak

\section{Results and discussion}

In spite of their different mathematical structure, both models yield very similar results. \linebreak
In the following, we will compare several quantities and comment on the likely causes of the remaining
differences between EST and PIC.

\begin{figure}
\center
\includegraphics[clip,width=0.8\linewidth]{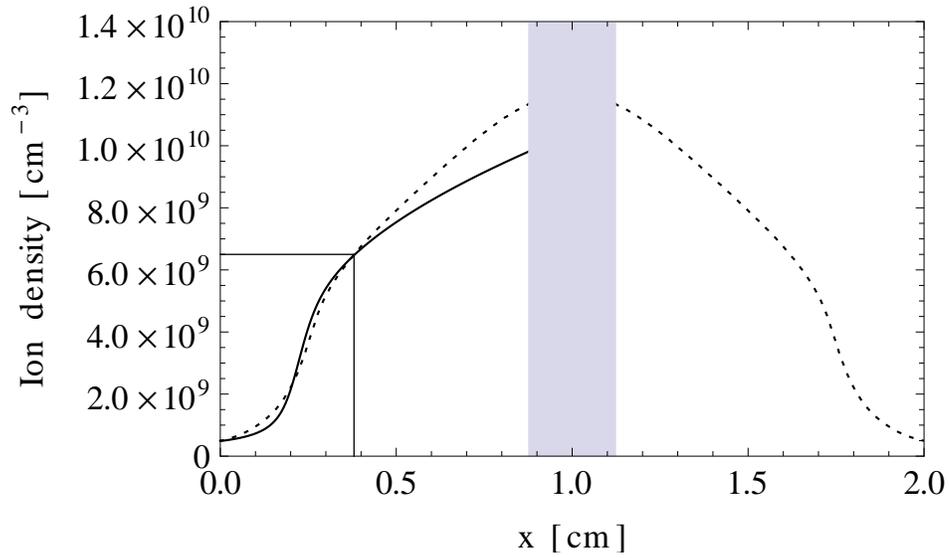}
\caption{The time averaged ion density in the discharge.
The dotted line represents the PIC/MCC results; the solid line those of EST.
The shaded area displays the area of the additional heating source. The rectangle in the left bottom of the figure denotes the maximal extension 
of the RF modulated sheath.}
\label{niESTPIC}
\end{figure}

Our first comparison  -- see \figurename{~\ref{niESTPIC}} -- is between the phase averaged ion densities. \linebreak
The largest deviation can be seen in the bulk, increasing towards the discharge center.
We believe that this deviation is due to the relatively low pressure, and it provides in fact an opportunity to illustrate 
one of the limitations of EST: For argon at $p= 1\,{\rm  Pa}$,
\linebreak
 the ion mean free path is $0.7\,{\rm cm}$. The PIC ions, which are mostly
born cold in the shaded ionization zone, travel only the same distance to the sheath edge. 
This does not suffice to establish drift equilibrium. The EST ions, on the other hand, are started at $x_{\rm B}$\linebreak already with drift speed
-- in the logic of EST, more specific (= non-fluiddynamic) information is not available.  
In other words: For $\lambda \approx L,$ where (1) is marginally violated,\linebreak also the bulk exhibits kinetic 
effects which are not captured by EST.

\begin{figure}
\center
\includegraphics[clip,width=0.8\linewidth]{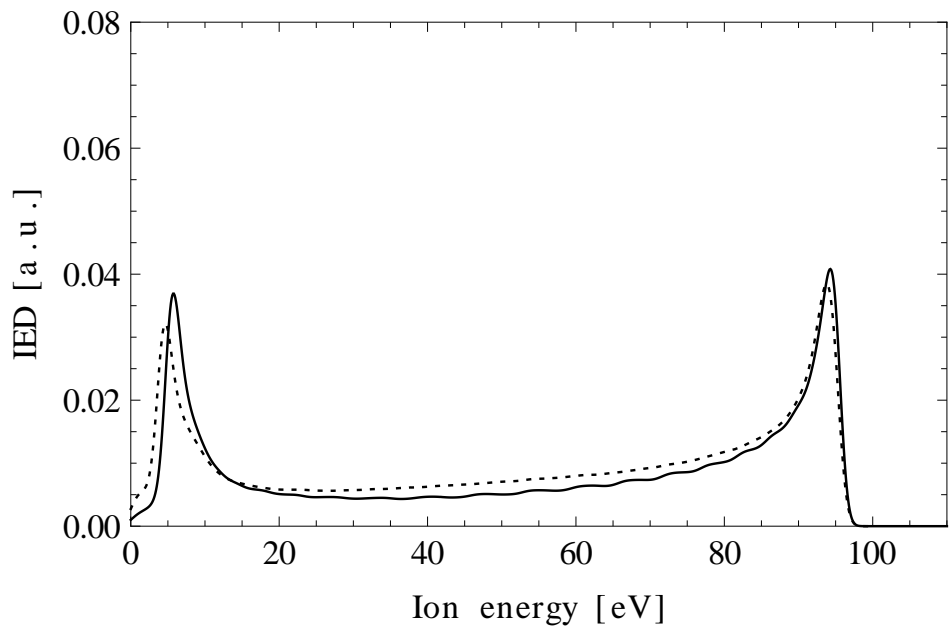}
\caption{Ion energy distribution function (IEDF) at the electrode. The dotted line represents the PIC/MCC results, the solid line those of EST.}
\label{IEDESTPIC}
\end{figure}

\begin{figure}
\center
\includegraphics[clip,width=0.8\linewidth]{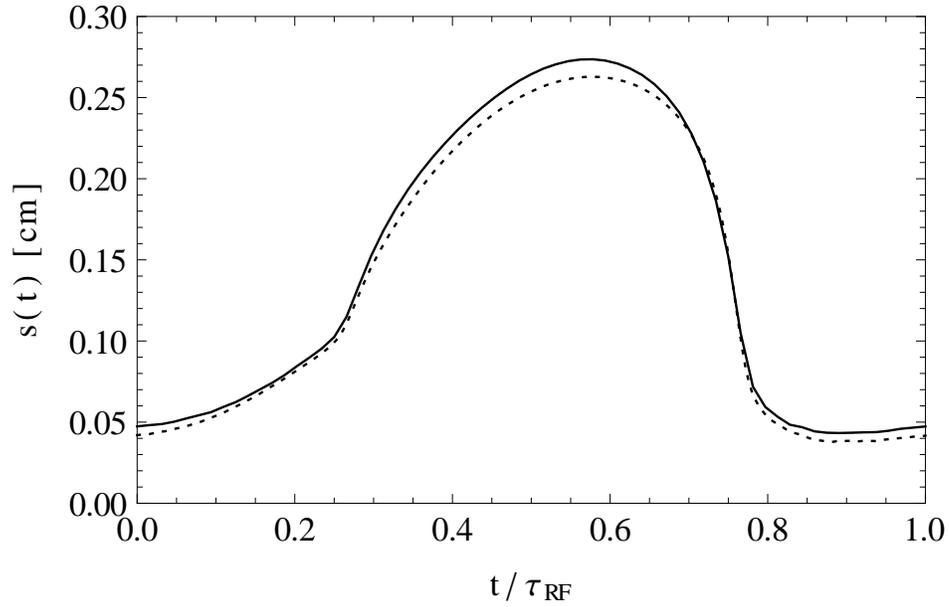}
\caption{The momentaneous sheath width as a function of normalized time $t/\tau_{\rm RF}$.
The dotted line represents the PIC/MCC results, the solid line those of EST.}
\label{SW}
\end{figure}

\begin{figure}
\center
\includegraphics[clip,width=0.8\linewidth]{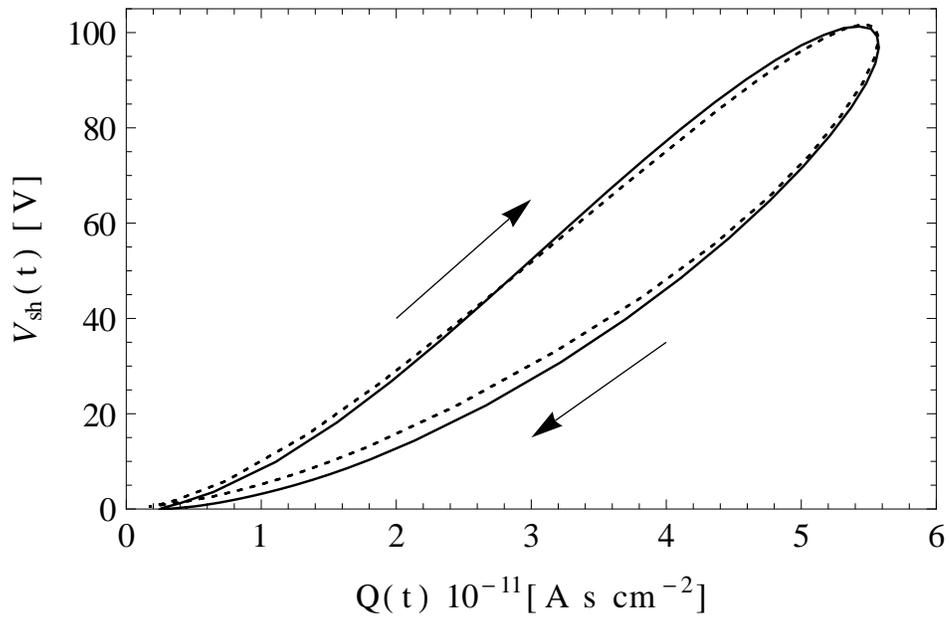}
\caption{Parametric plot of the sheath potential $V$ and the sheath charge per area $Q$ \linebreak as a function of time $t$.
         The arrows indicate the orientation.
The dotted line represents the PIC/MCC results, the solid line those of EST.}
\label{VSHQ}
\end{figure}

In the sheath itself, however, the named effect does not play a role, as the potential differences are 
much larger and both PIC and EST ions are
far from drift equilibrium. The initial conditions matter less, and the agreement of the ion densities is much better.
Any remaining discrepancies are probably 
due to small differences in the treatment of collisions, and due to the fact that the EST scheme assumes the  
electrons to be in Boltzmann equilibrium, while the PIC method treats them kinetically. 
(There may be a third possible cause for differences: EST uses a noise-reducing Fourier scheme to reconstruct the ion densities 
from the trajectories through the whole space-time, while PIC has only access to the instantaneous state and cannot remove any noise. 
However, this effect is not very influential.) 
The good agreement carries over to other sheath quantities:
\figurename{~\ref{IEDESTPIC}} shows the ion energy distributions at the electrode, \figurename{~\ref{SW}} the phase resolved sheath thickness $s(t)$, 
calculated using the definition of \cite{Brinkmann2009}, and \Fref{VSHQ} the charge-voltage characteristics. 
The results of EST (solid line) and of PIC/MCC (dotted line) are nearly identical. 

\begin{figure}
\center
\includegraphics[clip,width=0.8\linewidth]{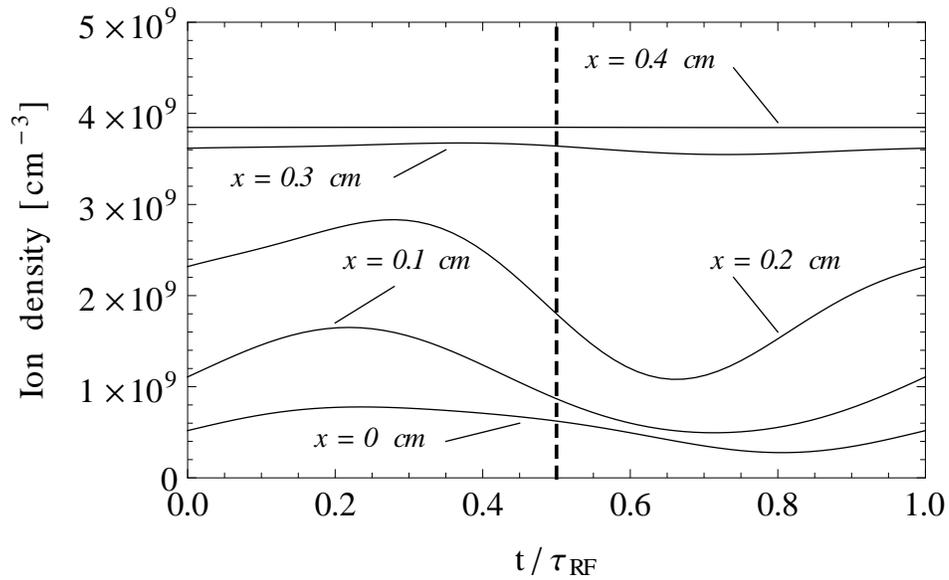}
\caption{The ion density $n_{\rm i}$ as a function of the normalized time $t/\tau_{\rm RF}$ at different positions $x$ in the simulation domain.}
\label{ni}
\end{figure}

As discussed in the introduction, it is in particular the faithful representation of the ion energy distribution at the electrode
that legitimates EST as a post-processor.  Physically, however, the charge voltage characteristics of \Fref{VSHQ} and its 
hysteresis is even more interesting: It demonstrates that both models capture kinetic effects correctly.
To understand the argument, consider the space-time trajectories of an ensemble of ions as displayed in \Fref{X}. 
Here, for simplicity, thermal spread -- ion temperature -- is neglected and ion-neutral collisions are turned off. 
The ions enter the depicted interval with Bohm speed and with a uniform phase distribution. As long as they are outside the sheath, 
they experience no electric field and are not accelerated. This motion translates into a temporally constant ion density, see \Fref{ni}. 
However, once the ions cross the momentaneous sheath edge $s(t)$,
they get accelerated and quickly drawn to the electrode. 
For $\omega_{\rm RF}\approx \omega_{\rm pi}$, this is clearly not a temporally symmetric effect, as more ions are collected when 
the sheath expands, as when it retracts. The ion flux in the sheath is thus temporally modulated, and so is the
ion density $n_{\rm i} (x,t)$ see  \Fref{ni}. The resulting global effect is the phase shift between the 
the sheath voltage $V(t)$ and the sheath charge $Q(t)$ that was depicted in \Fref{VSHQ}. Clearly, the
effect vanishes both for $\omega_{\rm RF}\ll \omega_{\rm pi}$ and $\omega_{\rm RF}\gg \omega_{\rm pi}$:
In the first case, the
sheath expansion is slow compared to the ion speed and the sheath-entering flux is independent of phase.
In the second regime, \linebreak the field changes too quickly for the ions to follow, and only the phase-averaged field 
determines the motion. 

The temporal asymmetry of the sheath charge-voltage relation surely impacts the overall behavior of a discharge. 
It will be interesting  to reconsider the studies on the plasma series resonance \cite{Mussenbrock2006,Czarnetzki,Bora,Schultze2010PSR} and stochastic 
heating \cite{Mussenbrock2007,Schulze2008,Ziegler2009} in the regime $\omega_{\rm RF}\approx \omega_{\rm pi}$ where a full 
$V$-$Q$-hysteresis is present.

\section{Conclusion}

Two kinetic models have been compared in the intermediate radio frequency regime, Particle-in-Cell/Monte Carlo collisions (PIC/MCC) and 
Ensemble-in-Spacetime (EST). \linebreak
The EST model, resolving the sheath alone, yields ion energy distributions and charge-voltage-relations close to those obtained 
by the fully kinetic PIC/MCC simulations, \linebreak provided that the flux of the
ions into the sheath, the sheath voltage, and the average electron temperature are known.
It can, therefore, be used as a post-processor to restore kinetic information from the limited
 information provided by fluid plasma models. \linebreak The gain in efficiency is considerable.
However, the PIC/MCC method remains the standard against which all other schemes must be compared.


\section{Acknowledgments}

This work has been supported by the Deutsche Forschungsgemeinschaft DFG via the collaborative research center TRR 87, and by the Hungarian Fund for scientific Research, grant K7753 and  K 105476.


\pagebreak

\newpage

\newpage

\newpage


\newpage

\vfill

\newpage

\vfill

\newpage

\end{document}